\documentclass[twocolumn,english,showpacs,preprintnumbers,prb]{revtex4-1}
\usepackage[latin9]{inputenc}
\usepackage[english]{babel}
\usepackage{setspace}
\usepackage{graphicx}
\usepackage{amssymb}
\usepackage{amscd}
\usepackage{rotating}
\usepackage{color}
\usepackage{epstopdf}
\usepackage{comment}

\usepackage{float}
\usepackage{natbib}
\usepackage{hyperref}
\usepackage[normalem]{ulem}%To strikeout the text

\newcommand{\beq}{\begin{equation}}
\newcommand{\eeq}{\end{equation}}

\newcommand{\barr}{\begin{eqnarray}}
\newcommand{\earr}{\end{eqnarray}}

\begin{document}

\title{Isolating Majorana fermions with finite Kitaev nanowires and temperature: the universality of the zero-bias conductance}

\author{V. L. Campo Jr$^1$, L. S. Ricco$^2$ and A. C. Seridonio$^{2,3}$}

\affiliation{$\mbox{}^1$Departamento de F\'isica, Universidade Federal de S\~ao Carlos, Rodovia Washington Luiz, km 235, Caixa Postal 676, 13565-905, S\~ao Carlos, S\~ao Paulo, Brazil}

\affiliation{$\mbox{}^2$Departamento de F\'isica e Qu\'imica, Unesp - Univ Estadual Paulista, 15385-000, Ilha Solteira, S\~ao Paulo, Brazil}

\affiliation{$\mbox{}^3$Instituto de Geoci\^encias e Ci\^encias Exatas - IGCE, Universidade
Estadual Paulista, Departamento de F\'isica, 13506-970, Rio Claro, S\~ao
Paulo, Brazil}

%\date{\today}

\begin{abstract}
 The zero-bias peak (ZBP) is understood as the definite signature of a Majorana bound state (MBS) when attached to a semi-infinite Kitaev nanowire (KNW) nearby zero temperature. However, such characteristics concerning the realization of the KNW constitute a profound experimental challenge. We explore theoretically a QD connected to a topological KNW of finite size at non-zero temperatures and show that overlapped MBSs of the wire edges can become effectively decoupled from each other and the characteristic ZBP can be fully recovered if one tunes the system into the leaked Majorana fermion fixed point. At very low temperatures, the MBSs become strongly coupled similarly to what happens in the Kondo effect. We derive universal features of the conductance as a function of the temperature and the relevant crossover temperatures. Our findings offer additional guides to identify signatures of MBSs in solid state setups.
\end{abstract}
\maketitle

\textit{Introduction.}---After the advent and understanding of topological
phases of matter, the proposal of decoherence-free topological quantum
computation~\cite{kitaev2003fault,sarma2015majorana,nayak2008non}, including
operations with isolated Majorana quasiparticle excitations, has triggered a
remarkable theoretical and experimental synergy in the condensed matter
physics community ~\cite{alicea2011non,leijnse2012introduction,alicea2012new,beenakker2013search}. Among the several theoretical proposals
~\cite{fu2008superconducting,sau2010generic,alicea2010majorana,lutchyn2010majorana,oreg2010helical,linder2010unconventional,cook2011majorana,nadj-perge2013proposal}, the one-dimensional topological Kitaev nanowire (KNW), exhibiting
\textit{p-}wave superconductivity~\cite{kitaev2001unpaired} has been considered
the paramount candidate to engineer isolated Majorana bound states (MBSs)
at its ends.

The presence of the isolated MBSs at the edges of the KNW is inferred from
tunneling spectroscopy, by analyzing the behavior of the zero-bias peak (ZBP)
in the conductance profiles~\cite{mourik2012signatures,das2012zero,deng2012anomalous,churchill2013superconductor,finck2013anomalous,deng2014parity,magchain,higginbotham2015parity,zhang2016ballistic,wire2016}, which should provide a
hallmark of the MBS presence. This demands manufacturing long KNWs to prevent
the MBSs overlapping and the consequent ZBP quenching at very low
temperatures, what is considered a hard experimental challenge~\cite{mourik2012signatures,das2012zero,deng2012anomalous,churchill2013superconductor,finck2013anomalous,deng2014parity,magchain,higginbotham2015parity,zhang2016ballistic,wire2016}.

\begin{figure}[t]
\centerline{\includegraphics[width=3.5in,keepaspectratio]{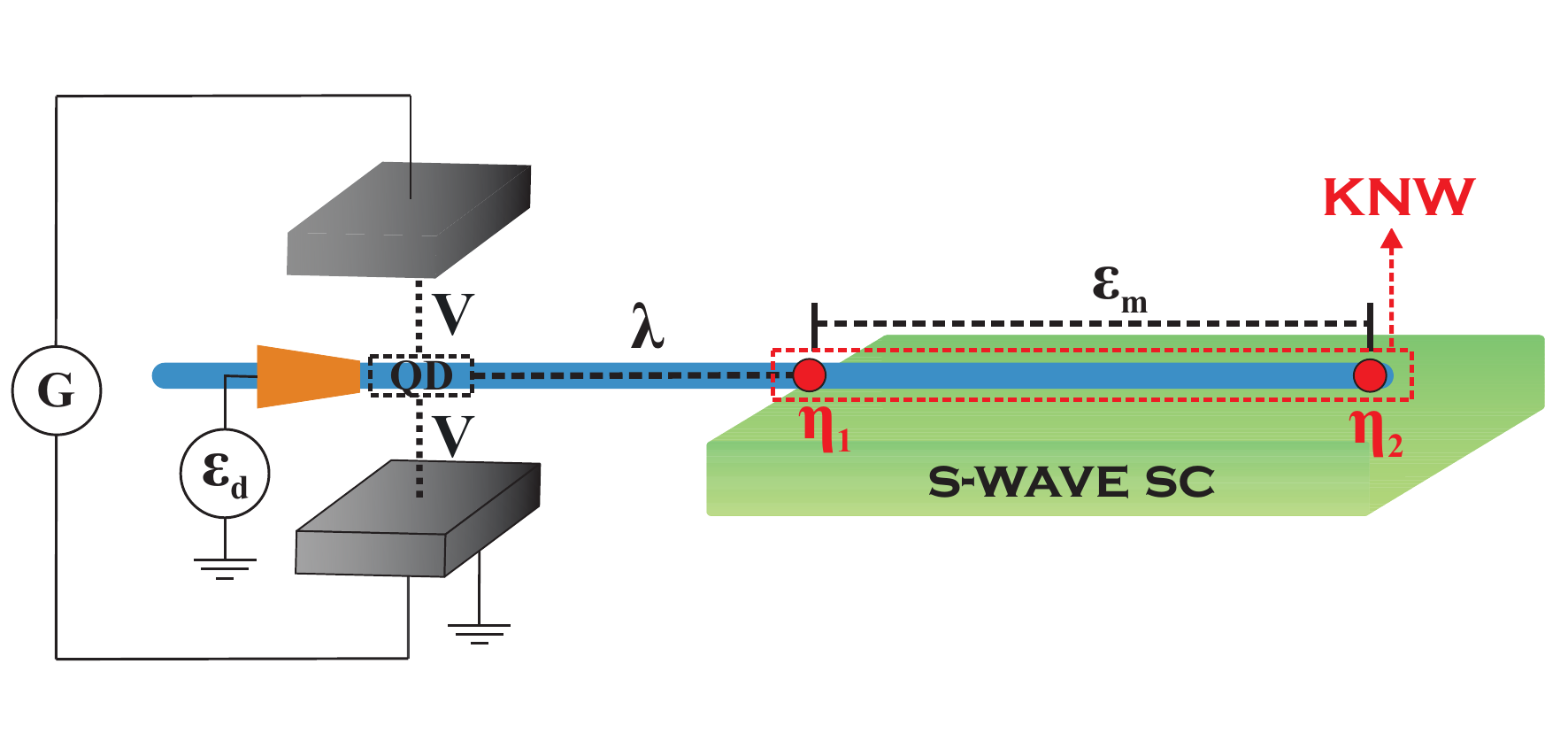}}
\caption{Sketch of the experimental setup, based on the recent experiment
  performed by M.T.Deng \textit{et al.}~\cite{wire2016}. A piece of
  semiconductor nanowire is placed on a s-wave superconductor (SC) material.
  In this semiconductor-superconductor segment, by considering suitable Zeeman
  field and spin-orbit coupling, a topological KNW emerges, giving rise to
  overlapped ($\varepsilon_{m}$) MBSs $\eta_{1}$ and $\eta_{2}$ at their edges.
  A QD (gate-tunable energy $\epsilon_{d}$) between two metallic leads
  (coupling $V$) is built in the semiconducting segment, where $\eta_{1}$
  leaks into (coupling $\lambda$) and can be detected as a ZBP.
\label{Setup}}
\end{figure}

In this work, we explore the quantum dot (QD)-Kitaev nanowire (KNW) hybrid
setup~\cite{liu2011detecting,cao2012probing,lee2013kondo,vernek2014subtle,lopez2014thermoelectrical,leijnse2014thermoelectric,liu2015probing,li2015probing,ruiz2015interaction} sketched in Fig.~\ref{Setup} based on
the recent experimental advances achieved by Deng \textit{et al}
~\cite{wire2016} in verifying the leakage of the MBS zero-mode into the QD,
which was first predicted theoretically in Ref.~[\onlinecite{vernek2014subtle}] by one of us. Such a scheme allows us to
probe the presence of the MBS by means of the zero-bias conductance sensing.
To shed light onto the large size problem stated above, we
considered the interplay between thermal broadening and overlapped MBSs
and found that effectively uncoupled edge-MBSs can pop-up at relatively large
temperature ranges.

We identify the fixed points of the model and perform a
renormalization group analysis
~\cite{yoshida2009universal,seridonio2009universal} to study the crossovers
between them and the temperature dependence of the conductance. The leaked
Majorana fixed point (LM), accounting for the leakage process
~\cite{wire2016,vernek2014subtle}, is seen to occur in the vicinity of a
characteristic temperature that depends solely on the KNW properties,
although the vicinity width depends on the whole set of model parameters.
Further, we find rigorously the crossover temperatures and derive an
analytic expression describing the universal behavior of the zero-bias
conductance along the crossovers. As it happens in the Kondo effect
~\cite{yoshida2009universal,seridonio2009universal}, the universal behavior
reveals a more complete signature of the physical system.

\textit{Model and fixed points.}---Assuming that the Zeeman field is large
enough in the system, so that we can neglect the transport of
spin down electrons through it, we consider the effective model with spinless
fermions~\cite{liu2011detecting,ruiz2015interaction}, whose Hamiltonian is
given by
\barr
H &=& \sum_{k,\alpha=U,L} \epsilon_{k} c_{k,\alpha}^\dagger c_{k,\alpha} +
V\sum_{k,\alpha=U,L}\left(c_{k,\alpha}^\dagger d + d^\dagger c_{k,\alpha} \right)\nonumber\\
&+& \epsilon_d d^\dagger d + i\epsilon_m \eta_1\eta_2 + \sqrt{2}\lambda(d - d^\dagger)\eta_1,\label{Ham1}
\earr
where the first term describes the conduction electrons in the upper (U) and
lower (L) leads. We assume half-filled conduction bands in the particle-hole
symmetric regime, with a constant density of states equal to $\rho$,
$-D\le\epsilon_{k,\alpha}\le D$ and Fermi energy equal to zero. The QD here has
only one energy state $\epsilon_d$ that is hybridized with the conduction
states in the leads through the second term in the Hamiltonian, resulting in a
linewidth $\Gamma = \pi\rho V^2$. We assume here symmetric coupling to the
leads. The KNW is assumed to be in the topological phase with two MBSs at its
ends ($\eta_i = \eta_i^\dagger$, $\eta_i\eta_i = 1/2$), with an overlap amplitude $\epsilon_m \sim e^{-L/\xi}$  between them, where $L$ is the length of the KNW and $\xi$ is the superconductor coherence length. The last term in the Hamiltonian represents the coupling between the MBS $\eta_1$ and the QD single state.

We consider now even and odd conduction states, $e_k = (c_{k,U}+c_{k,L})/\sqrt{2}$ and $o_k = (c_{k,U}-c_{k,L})/\sqrt{2}$ and also the nonlocal fermionic operators $b = (\eta_1 + i\eta_2)/\sqrt{2}$ and $b^\dagger = (\eta_1 - i\eta_2)/\sqrt{2}$ ($\left\{b,b^\dagger\right\}=1, \left\{b,b\right\}=0$), to rewrite the model Hamiltonian as
\barr
H &=& \sum_{k} \epsilon_{k} \left(e_{k}^\dagger e_{k} +  o_{k}^\dagger o_{k} \right)
+ \sqrt{2}V\sum_{k} \left(e_{k}^\dagger d + d^\dagger e_{k} \right) + \epsilon_d d^\dagger d\nonumber\\
&+& \epsilon_m \left(b^\dagger b - \frac{1}{2}\right) - \lambda(d^\dagger b + b^\dagger d + d^\dagger b^\dagger + b d), \label{Ham2}
\earr
where the odd conduction states are decoupled from the QD and the number of fermions is not conserved.

The zero-bias conductance as a function of the temperature $T$ can be calculated from\cite{yoshida2009universal}
\beq
G(T) = \frac{2e^2}{h}\pi \Gamma \left[ \frac{1}{k_BT}\frac{1}{Z}\sum_{n,m}\frac{|\langle n |d| m \rangle|^2}{e^{\beta E_n} + e^{\beta E_m}}\right]\label{Gcalc1}
\eeq
or, alternatively, we can also rewrite Eq.~(\ref{Gcalc1}) as~\cite{Jauho}
\beq
G(T) = \frac{2e^2}{h} \Gamma \int_{-\infty}^\infty\:\text{{Im}}\{G_{d,d}(\omega)\} \left(\frac{\partial f}{\partial \omega}\right)\:d\omega,\label{Gcalc2}
\eeq
where $f(\omega)$ is the Fermi-Dirac function. The QD Green's function can be promptly obtained from the equation of motion~\cite{Jauho} procedure, leading to
\beq
G_{d,d}(\omega) = \frac{C(\omega)}{A(\omega)C(\omega) - B^2(\omega)},\label{greendd}
\eeq
where $A(\omega)=\omega - \epsilon_d + 2i\Gamma - B,$ $C(\omega)=\omega + \epsilon_d + 2i\Gamma - B$ and
\barr
B(\omega) &=& \frac{2\lambda^2\omega}{\omega^2 - \epsilon_m^2},\label{Bomega}
\earr
that reproduces the well-known Green's function for a QD side-coupled to a
topological KNW found in Ref.[\onlinecite{liu2011detecting}], here expressed
differently for convenience once we target to show the system universality.

\begin{figure}[t]
\centerline{\includegraphics[width=3.4in,keepaspectratio]{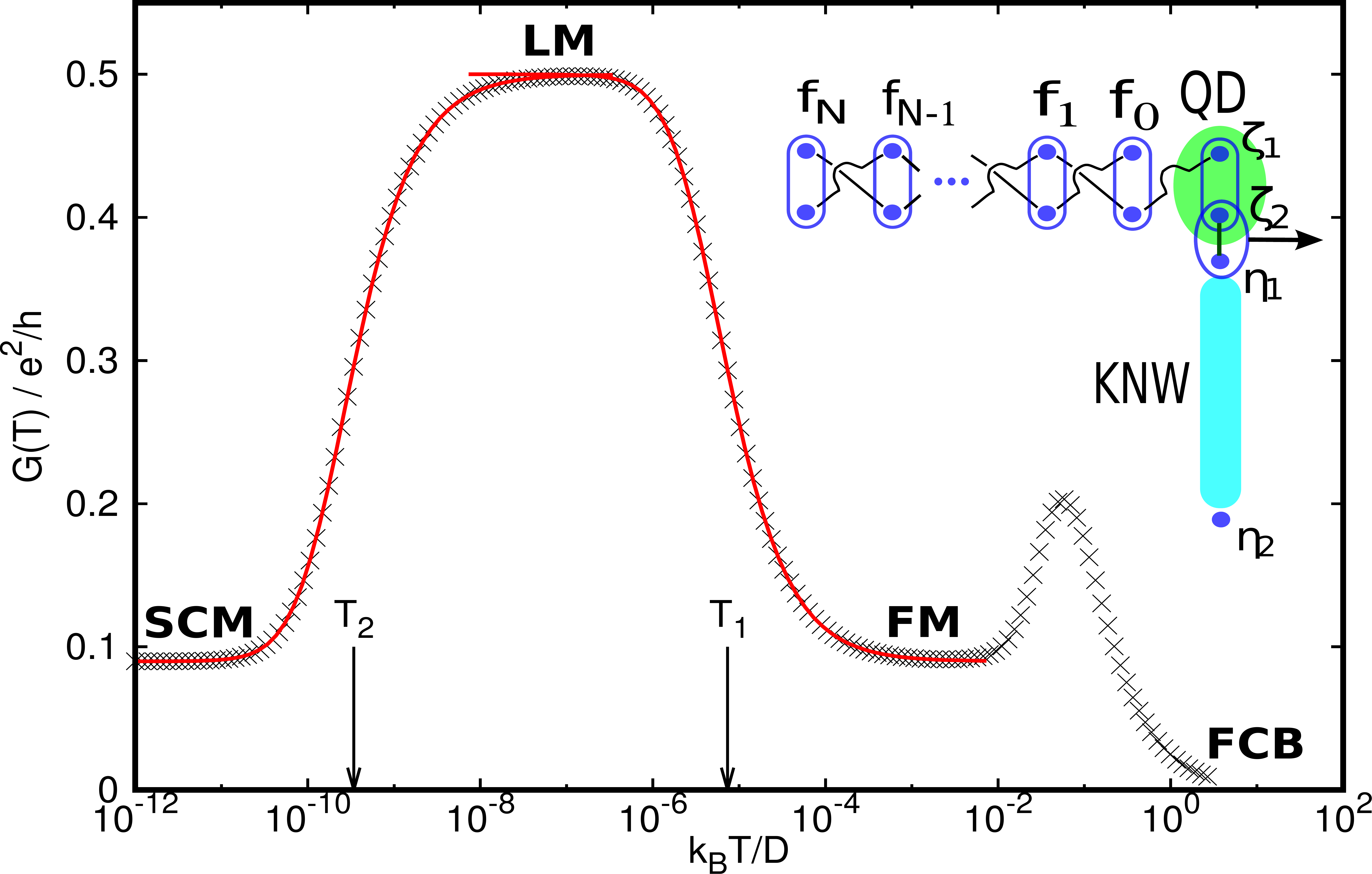}}
\caption{Zero-bias conductance as a function of the
  temperature in logarithmic scale for $\epsilon_d = 0.1D$,
  $\Gamma = 0.005\pi D$, $\lambda = 0.0008D$, $\epsilon_m = 5\times10^{-8}D$.
  The model parameters were carefully chosen to make the several fixed points
  very clear. Continuous lines are given by Eqs.~(\ref{Guniv0}) and
  (\ref{Huniv}) with crossover temperatures $T_1$ (Eq.~(\ref{T1})) and
  $T_2$ (Eq.~(\ref{T2})).
  The inset pictorially shows the Majorana representation of the problem in
  the leaked Majorana (LM) fixed point, considering a tight-binding
  description of the conduction band, comprising two uncoupled zig-zag chains
  of Majorana fermions (blue circles). Majorana fermions $\eta_1$ and
  $\zeta_2$ are strongly coupled and removed from the system (see the
  horizontal arrow). One of the Majorana fermion chains gives rise to half
  of the single-particle excitations of the free conduction band, while the
  other chain, coupled to the Majorana fermion $\zeta_1$, gives rise to half
  of the single-particle excitations of the $\epsilon_d =0$ resonant level
  model. This implies a conductance equal to half of that in such resonant
  level model, explaining the characteristic value 0.5~$e^2/h$ in the LM
  fixed point.
  \label{fig2}}
\end{figure}

For a vanishing coupling $\lambda$ between the QD and the KNW or for
$\epsilon_m \to \infty$, so that the fermionic state $b$ becomes empty, we end
up with a simple resonant level model. For non-vanishing coupling $\lambda$
and $\epsilon_m=0$ (infinite KNW), we can find from
Eqs.~(\ref{Gcalc2}) and (\ref{greendd}) that the zero-bias conductance through
the QD approaches 0.5~$e^2/h$ at $\omega=0$, when the temperature $T\to0$, whatever the values of $\epsilon_d$ and $\Gamma$, being a signature of the leakage of the MBS $\eta_{1}$ into the QD\cite{vernek2014subtle}. The problem is that any nonzero $\epsilon_m$ makes the conductance change to its resonant level model value
\beq
G_0 = \frac{e^2}{h} \frac{4\Gamma^2}{\epsilon_d^2 + 4\Gamma^2} = \frac{e^2}{h}~\sin^2(\delta)
\eeq
in the limit of zero temperature, where $\delta$ is the phase-shift at the
Fermi level. This fact prompt us to a more detailed investigation of the
temperature dependence of the conductance.

\textit{Results and Discussion.}--- Fig.~\ref{fig2} is clarifying. At high
temperatures, when we can effectively consider both $\lambda$ and $\epsilon_m$
equal to zero, the conductance approaches $G_0 \approx 0.09$~$e^2/h$. Then, as
the temperature is lowered, the coupling finally emerges, leading to a
crossover from a conductance equal to $G_0$ towards 0.5~$e^2/h$. This value
remains stable in a certain temperature range. At some point, however, the
tiny energy $\epsilon_m$ dominates and the MBSs become strongly coupled,
yielding a new crossover ending with $G_0$ recovered. This
final crossover resembles the Kondo
effect~\cite{yoshida2009universal,seridonio2009universal},
where a localized spin is screened by the conduction electrons in spite of how
weak can be the coupling between them. If the coupling is very small, the
crossover will occur at a extremely low Kondo temperature and eventually can
become unobserved. Analogously, in our case, if the KNW is long enough
$(\epsilon_{m}\to0)$, the last crossover can be shifted to very low temperatures,
allowing the observation of an essentially stable conductance value equal to
0.5~$e^2/h$. As it happens in the Kondo effect, more relevant than particular
values of some physical properties in the $T\to 0$ limit is the universal
behavior of these physical properties during the crossover as some parameter
is changed, typically the temperature. Below, we recognize
explicitly the accounted fixed points and then proceed to the analysis
concerning the temperature dependence of the conductance.

\textit{Free conduction band (FCB) fixed point}. This corresponds to do
$V=\epsilon_d=\lambda=\epsilon_m= 0$ in the model Hamiltonian. Both QD level
and nonlocal fermionic level $b$ are detached from the conduction band
and can be empty or occupied, so that any energy is fourfold degenerate. The
excitations are those in a free conduction band. The system would be close to
this fixed point at temperatures
$k_BT \gg \Gamma,\epsilon_d,\lambda,\epsilon_m$, where the conductance goes to
zero. Since the temperature must necessarily be lower than the effective
superconducting gap in the KNW, this fixed point will not be observed in
general.

\textit{Free Majorana fermions (FM) fixed point}. In this
case, $\lambda=\epsilon_m = 0$ and the resulting resonant level model must
be considered in the limit $k_BT \ll \Gamma$. The energies are twofold
degenerate, since the nonlocal fermionic level $b$ can be empty or occupied.
The excitations in the conduction band have a phase-shift $\delta$, with
$\cot\delta = \epsilon_d/2\Gamma$.

\textit{Strongly coupled Majorana fermions (SCM) fixed point}. Here, we regard
$\epsilon_m\to\infty$. The MBSs become strongly coupled and the nonlocal
fermionic level $b$ remains empty. The system becomes the resonant level model,
with the same conductance and same excitations as in the FM fixed point, but
without degeneracy.

\textit{Leaked Majorana fermion (LM) fixed point:} This corresponds to do
$\lambda\to\infty$ in the model Hamiltonian. From Eq.~(\ref{Ham1}), the MBSs
$\eta_1$ and $\zeta_2 = i(d^\dagger - d)/\sqrt{2}$ become infinitely coupled,
leading to a nonlocal fermionic level with infinite energy, that remains
empty. But, we still have the MBS $\zeta_1 = (d^\dagger + d)/\sqrt{2}$  in the
QD, so that the leaked Majorana fixed point is described by the following
Hamiltonian:
\barr
H_{LM} &=& \sum_{k} \epsilon_{k} e_{k}^\dagger e_{k}
+ V\sum_{k} \left(e_{k}^\dagger \zeta_1 + \zeta_1 e_{k} \right).\label{HLM1}
\earr

Essentially, the MBS has leaked from the KNW edge into the
QD~\cite{vernek2014subtle,wire2016}. However, it is coupled to the even
conduction states so that this leaking process will reach the conduction band.

With the MBS $\zeta_1$ at the QD level and $\eta_2$ at the other far edge of
the KNW, we introduce a new fermionic operator,
$a = (\zeta_1 + i\eta_2)/\sqrt{2}$, to rewrite $H_{LM}$ as
\beq
H_{LM} = \sum_{k} \epsilon_{k} e_{k}^\dagger e_{k}
+ \frac{V}{\sqrt{2}}\sum_{k} \left(e_{k}^\dagger a + e_{k}^\dagger a^\dagger + \mathrm{h.c.} \right).\label{HLM2}
\eeq

As discussed in the caption of Fig.~\ref{fig2} and explained in detail in the
supplemental material, one half of the single-particle excitations of $H_{LM}$
are of free-conduction band type and another half of them are of $\epsilon_d=0$
resonant level model type. Only the last set of excitations can contribute to
the conductance, leading to the characteristic value of 0.5~$e^2/h$ as $T\to0$.

\begin{figure}[t]
\centerline{\includegraphics[width=3.4in,height=2.2in]{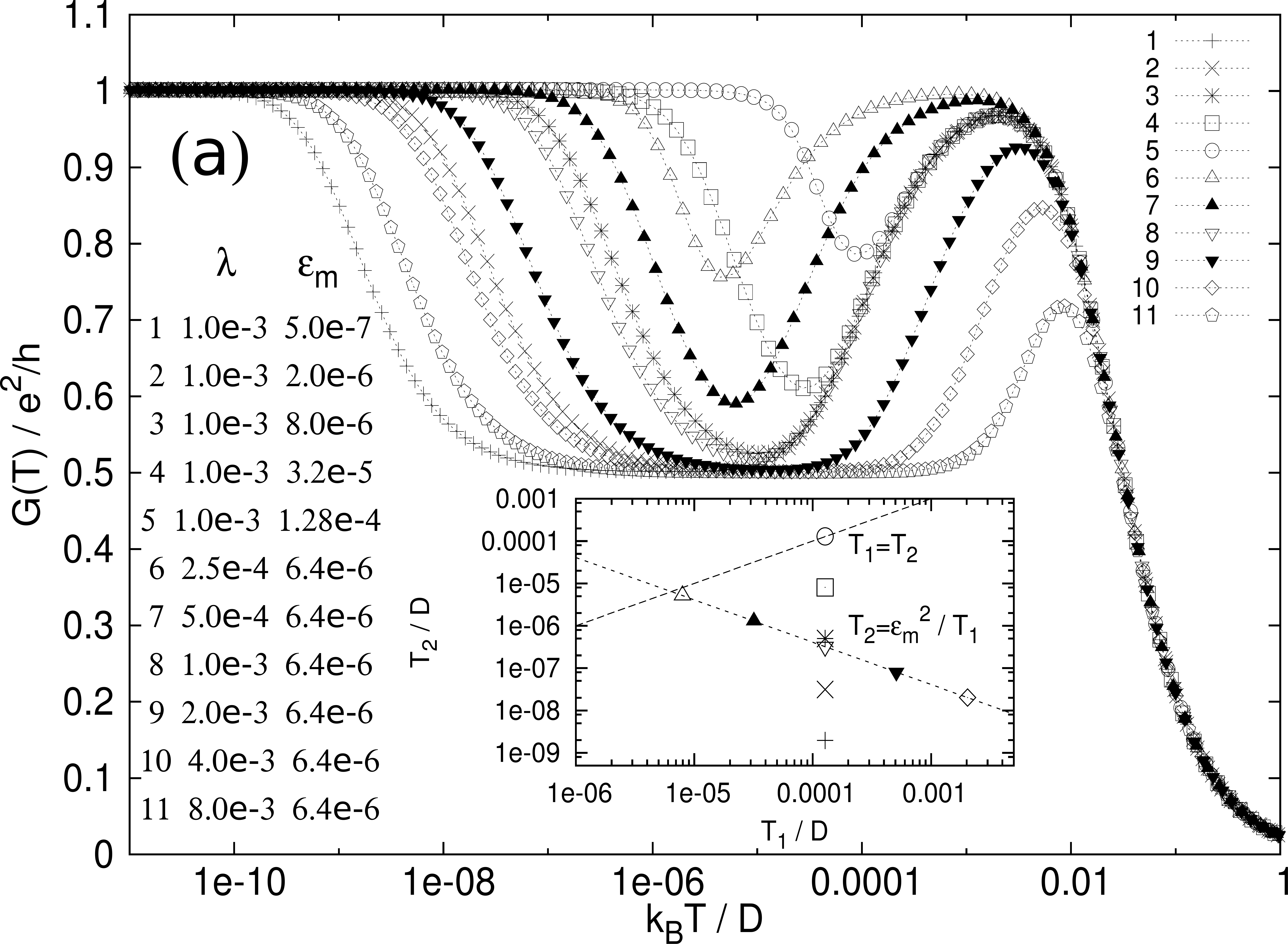}}
\centerline{\includegraphics[width=3.4in,height=2.2in]{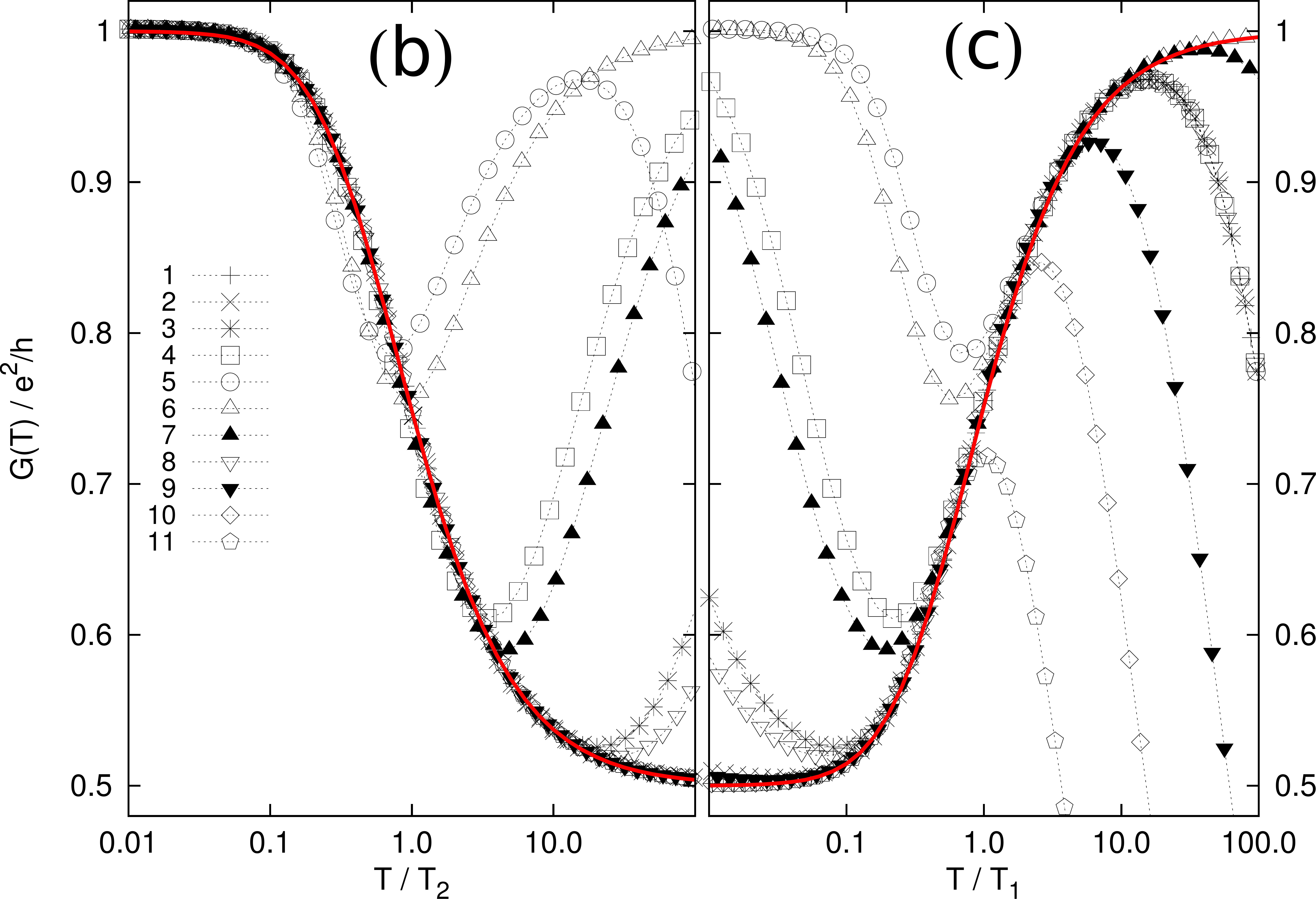}}
\caption{(a) Zero-bias conductance as a function of the temperature in
  logarithmic scale for 11 different model parameter pairs
  $(\lambda,\epsilon_m)$. In all curves, $\epsilon_d = 0$ and
  $\Gamma = \pi0.005~D$. The inset shows the distribution of the corresponding
  crossover temperatures $T_1$ and $T_2$ (Eqs.~(\ref{T1}) and (\ref{T2})).
  Curves 1-5 have the same $\lambda$ and the same $T_1$. As $T_2$ increases
  (with $\epsilon_m^2$), approaching $T_1$, the crossovers will merge at some
  point, and the LM fixed point ceases to be achieved. Curves 6-11 have the
  same $\epsilon_m$ and their minima between the crossovers are approximately
  at the same point (Eq.~(\ref{T1T2})). As $\lambda$ increases, $T_1$ rises
  and $T_2$ lowers, broadening the LM fixed point plateau. (b) and (c)
  Conductance as a function of $T/T_2$ and $T/T_1$, respectively. The points
  collapse into the continuous lines in Eq.~(\ref{Guniv0}) with
  ${\mathcal{H}}(t)$ in Eq.~(\ref{Huniv}). In (b), when $T_1$ is not much
  higher than $T_2$, deviations from the universal curve will start at low
  temperatures. In (c), deviations at $T<T_1$ occur when $T_2$ is not much
  lower than $T_1$ while deviations at $T>T_1$ occur when $T_1$ is not much
  smaller than $\Gamma/k_B$.
\label{fig3}}
\end{figure}

Now we turn to the problem of carefully identifying universal behavior in the
zero-bias conductance. In Fig.~\ref{fig3}a, we show the conductance for
different sets of model parameters.  We have used $\epsilon_d=0$ and
$\Gamma = \pi0.005~D$, changing $\epsilon_m$ and $\lambda$. Therefore, we have
conductance $G_0 =1.0~e^2/h$ in the FM and SCM fixed points. It is clear from
Fig.~\ref{fig3}a that the crossovers occur around parameter-dependent
temperatures $T_1$ (between the FM fixed point and the LM fixed point) and
$T_2$ (between the LM fixed point and the SCM fixed point). We have found
from the numerical results that
\beq
k_BT_1 = \frac{2}{\Gamma}\frac{\lambda^2}{\left[1 + \left(\frac{\epsilon_d}{2\Gamma}\right)^2\right]},\label{T1}
\eeq
and
\beq
k_BT_2 = \frac{\Gamma}{2}\left[1 + \left(\frac{\epsilon_d}{2\Gamma}\right)^2\right]\left(\frac{\epsilon_m}{\lambda}\right)^2. \label{T2}
\eeq

%It must be noted that $T_1 \propto \lambda^2$ not $\lambda$ as one could naively suggest. In the same way, $T_2 \propto \epsilon_m^2$ not $\epsilon_m$.

Physically, it is clear that $T_1$ must increase with $\lambda$ and that $T_2$
must increase with $\epsilon_m$. Differently from a naive expectation, we have
$T_2 \propto \epsilon_m^2$, not $\epsilon_m$. Except for the resonant case,
$\epsilon_d=0$, $T_1$ and $T_2$ are not monotonic functions of $\Gamma$. With
the remaining parameters fixed, when $\Gamma = |\epsilon_d|/2$, $T_1$ is
maximum and $T_2$ is minimum. For $\epsilon_d = 0$, increasing the
hybridization between the QD and the conduction band lowers $T_1$, since the
QD level mixes with the continuum of states, making the coupling with the
Majorana fermion $\eta_1$ less effective, and  increases $T_2$, once the
Majorana fermion that leaked to the conduction band around the LM fixed point
will couple to the Majorana fermion $\eta_2$ more easily with a strong
hybridization.

Scaling the temperature by $T_2$ or $T_1$, the crossover portions of the
different curves in Fig.~\ref{fig3}a collapse into the same curve as shown in
Figs.~\ref{fig3}b and \ref{fig3}c. In addition, we have
\beq
\sqrt{T_1T_2} = \epsilon_m/k_B,\label{T1T2}
\eeq
that means the plateau at 0.5$e^2/h$ in the conductance will
be centered at $T\sim\epsilon_m/k_B$, being well defined only if
$T_2 \ll \epsilon_m/k_B \ll T_1$. From Figs.~\ref{fig3}b and \ref{fig3}c,
we see that the temperature must be changed by at least two orders of
magnitude to complete the crossover, what demands $T_1 > 100T_2$ to clearly
have the system in the LM fixed point. The ratio
$T_1/T_2 = (k_BT_1/\epsilon_m)^2$ depends on all parameters, what helps to
tune it large enough.

In general, we expect~\cite{yoshida2009universal,seridonio2009universal} that
the conductance between a low-temperature fixed-point ($G=G_l$) and a
high-temperature fixed-point ($G=G_h$) be given by
\beq
G = \frac{G_l + G_h}{2} + \frac{G_l - G_h}{2}{\mathcal{H}}\left(\frac{T}{T^\star}\right),\label{Guniv0}
\eeq
where ${\mathcal{H}}(t)$ is a universal function
characteristic of the crossover and $T^\star$ is the crossover temperature. In
the present case, we have found that the same universal function
describes the crossover between SCM and LM fixed points and that between LM and FM fixed points.

In order to determine ${\mathcal{H}}(t)$ analytically, we consider the
SCM$\to$LM crossover, for instance, and assume $T_2 \ll \epsilon_m \ll T_1$ to
have the crossovers well separated. For temperatures
$T\sim T_2 \ll \epsilon_m/k_B$, an inspection in Eq.~(\ref{Gcalc2}) shows that
only energies $\omega \ll \epsilon_m$ are important, so we have
$B(\omega) \approx -2\left(\frac{\lambda}{\epsilon_m}\right)^2\omega$ in
Eq.~(\ref{Bomega}). Introducing $r = \frac{\lambda^2}{\epsilon_m^2}$,
$x = \frac{\epsilon_d}{2\Gamma}$ and $z = \frac{\omega}{k_BT_2}$, we have from
Eq.~(\ref{greendd}) that
\beq
G_{d,d} \approx -\frac{1}{4\Gamma(1+x^2)}\left[\frac{z(1+x^2) + 2x + i2}{1 - iz}\right], \label{simplerho}
\eeq
where we have exploited that $r\gg 1$ and that $2rk_BT_2 = \Gamma(1+x^2)$.
Substituting Eq.~(\ref{simplerho}) into Eq.~(\ref{Gcalc2}), defining $t = T /T_2$ and
changing to the variable $u = \omega/k_BT = z/t$, we get
\beq
\frac{G(T)}{e^2/h} = \int_{-\infty}^\infty \left[\frac{\frac{u^2t^2}{2} + \frac{G_0}{e^2/h}}{1+u^2t^2}\right]\frac{e^u}{\left(e^u + 1\right)^2}~du.\label{Guniv2}
\eeq
If we add and subtract $\frac{1}{2}\left(\frac{G_0}{e^2/h} + \frac{1}{2}\right)$ to the term between brackets in (\ref{Guniv2}) and use that $\int_{-\infty}^\infty \frac{e^u}{\left(e^u + 1\right)^2}du = 1$, we will finally obtain
\beq
\frac{G(T)}{e^2/h} = \frac{1}{2}\left(\frac{G_0}{e^2/h} + \frac{1}{2}\right) + \frac{1}{2}\left(\frac{G_0}{e^2/h} - \frac{1}{2}\right){\mathcal{H}}(t),\label{Guniv3}
\eeq
with the universal function given by
\beq
{\mathcal{H}}(t) = \int_{-\infty}^\infty \left[\frac{1 - u^2t^2}{1+u^2t^2}\right]\frac{e^u}{\left(e^u + 1\right)^2}~du.\label{Huniv}
\eeq

\textit{Conclusions.}---In summary, we have determined the universal behavior of the zero-bias conductance for the simple spinless model in Eq.~(\ref{Ham1}) along the crossovers connected to the LM fixed point. This enlarges the signature of the MBS in the end of the KNW and can help to reveal its presence when the LM fixed point is not fully achieved. Even with a finite KNW, it may be possible to set up the model parameters to have $T_1 \gg T_2$ and a reasonably large temperature range with conductance close to 0.5$e^2/h$. We expect that the current findings offer additional guides to identify signatures of MBSs in solid state setups.

\textit{Acknowledgments.}---We thank the funding Brazilian agencies CNPq (307573/2015-0), CAPES and S\~ao Paulo Research Foundation (FAPESP) - grant: 2015/23539-8.

\end{document}